\documentclass[aps,pre,preprint,showpacs,amsmath,amssymb,floatfix,tightenlines,12pt,superscriptaddress]{revtex4-1} % PRL -superscriptaddress

\usepackage[pdftex]{graphicx}       % Include figure files
\usepackage{txfonts}        % Times New Roman text & math font
\usepackage{amsmath}
\usepackage{amssymb}
\usepackage{mathrsfs}	% Extra script font
\usepackage{enumerate}

\newcommand{\scT}{\mathcal{T}}

\newcommand{\avm}{\langle m \rangle}
\newcommand{\avh}{\langle h \rangle}
\newcommand{\avz}{\langle z_0 \rangle}

\newcommand{\eref}[1]{Eq.~\eqref{#1}}
\newcommand{\fref}[1]{Fig.~\ref{#1}}
\newcommand{\sref}[1]{Section \ref{#1}}

\usepackage{hyperref} % this line at end of preamble
\hypersetup{colorlinks,citecolor=blue,filecolor=blue,linkcolor=blue,urlcolor=blue}

\begin{document}

\title{Monte Carlo simulation of 3-star lattice polymers pulled from an adsorbing surface}

\author{C. J. Bradly} 
\email{chris.bradly@unimelb.edu.au}
\affiliation{School of Mathematics and Statistics, University of Melbourne, Victoria 3010, Australia}
\author{A. L. Owczarek}
\email{owczarek@unimelb.edu.au}
\affiliation{School of Mathematics and Statistics, University of Melbourne, Victoria 3010, Australia}

\date{\today}

\begin{abstract}
	We study uniform 3-star polymers with one branch tethered to an attractive surface and another branch pulled by a force away from the surface.
	Each branch of the 3-star lattice is modelled as a self-avoiding walk on the simple cubic lattice with one endpoint of each branch joined at a common node.
	Recent theoretical work \cite{Rensburg2018} found four phases for this system: free, fully adsorbed, ballistic and mixed.
	The mixed phase occurs between the ballistic and fully adsorbed phase.
	We investigate this system by using the flatPERM  Monte Carlo algorithm with special restrictions on the endpoint moves to simulate 3-stars up to branch length 128.
	We provide numerical evidence of the four phases and in particular that the ballistic-mixed and adsorbed-mixed phase boundaries are first-order transitions.
	The position of the ballistic-mixed and adsorbed-mixed boundaries are found at the expected location in the asymptotic regime of large force and large surface-monomer interaction energy. 
	These results indicate that the flatPERM algorithm is suitable for simulating star lattice polymers and opens up new avenues for numerical study of non-linear lattice polymers.
\end{abstract}

\maketitle

%%%%%%%%%%%%%%%%%%%%%%%%%%%%%%%%%%%%%%%%%%%%%%%%%%%%%%%%%%%%%
%%%%%%%%%%%%%%%%%%%%%%%%%%%%%%%%%%%%%%%%%%%%%%%%%%%%%%%%%%%%%
\section{Introduction}
\label{sec:Intro}
The self-avoiding walk is the canonical model for linear polymers in a dilute solution \cite{Rensburg2015}.
Of the many variations to this model, the adsorption of polymers to a surface has attracted considerable attention \cite{Eisenriegler1982,Vrbova1999,Grassberger2005,Luo2008,Klushin2013,Bradly2018}.
For a walk with one end tethered to an attractive surface the configuration of the polymer is that of an extended coil dominated by entropic repulsion at high temperatures.
There is a critical temperature $T_\text{c}$ below which the polymer is adsorbed to the surface, in a configuration where large fractions of the monomers are attached to the surface \cite{DeBell1993}.
Studies of the adsorption of polymers have also considered non-linear polymers and such as branched polymers and lattice stars \cite{Ohno1991}.

A related variation considers the effect of forces applied to the polymer, motivated by advances in the application of atomic force microscopy to manipulate single polymers attached to a surface \cite{Haupt1999,Zhang2003}. 
The addition of a pulling force to induce desorption can be applied to models of the adsorption of lattice polymers in a range of different ways including self-avoiding walks \cite{Orlandini2016,Rensburg2013,Beaton2015,Mishra2005,Guttmann2014}, self-avoiding polygons \cite{Beaton2017,Beaton2015b,Guttmann2018} and at variable points along the walk \cite{Rensburg2017}. 
As well as linear polymers one can consider a large array of branched polymers including combs, spiders and stars \cite{Rensburg2018b}.
These studies have revealed new thermodynamic phases such as a ballistic phase where the polymer configuration depends only on the single constant pulling force as well as mixed phases between this and the fully adsorbed phase.

In this work we continue the study of uniform 3-star polymers in three dimensions, using Monte Carlo simulations to verify recent theoretical results \cite{Rensburg2018}.
Due to these results, the phase diagram in three-dimensional case is fairly well understood and it serves as a useful test case for numerical work before turning to more intractable problems in other dimensions or involving different classes of branched polymers.
However, even for the cubic lattice there remain questions about the nature and location of some of the phase transitions.

We view an $f$-star as a connected graph embedded in a lattice composed of $f$ self- and mutually-avoiding walks, called branches.
One endpoint of each branch occupies a common central node. 
For a uniform $f$-star, which we denote $\psi_n$, each branch has length $n$ steps for a total of $fn+1$ nodes.
The cases \smash{$f=1$} and \smash{$f=2$} reduce to the canonical self-avoiding walk (SAW) of length $n$ and $2n$, respectively.
One branch is tethered to an impermeable adsorbing surface and another branch is pulled by a constant force $F$. 
The $f$-star may only occupy lattice sites in one of the half-spaces defined by the surface, or on the surface itself. 
The surface-monomer interaction is modeled by assigning an energy $-\epsilon$ to all $m$ points contacting the surface, excepting the permanently tethered point.
The canonical model would be to place the tether point at the end of one of the branches, but in this work we consider the tether to be at an arbitrary point on one of the non-pulled branches.
This is due to constraints from the algorithm and we will see that it has no effect on the phase diagram.
The pulling is modeled by considering the work required to pull the end of one of the non-tethered branches of a fully adsorbed configuration away from the surface and up to a height $h$ relative to the surface.

The total energy of $\psi_n$ is \smash{$hF + m\epsilon$}, to which we assign the corresponding Boltzmann weight $a^m y^h$, where \smash{$a = \exp(\epsilon/k_\text{B}T)$} and \smash{$y = \exp(F/k_\text{B}T)$}. 
The partition function of the set $\scT_n$ of such $f$-stars is
\begin{equation}
    Z_n(a,y) = \sum_{\psi_n \in \scT_n} a^m y^h = \sum_{m,h} c_{n,m,h}^{(f)} a^m y^h,
    \label{eq:Partition}
\end{equation}
where $s_{n,m,h}^{(f)}$ is the number of $f$-stars with branch length $n$ confined to one side of an impermeable surface, with $m$ contacts with that surface and the endpoint of the pulled branch at height $h$.
Then the (reduced) free energy is
\begin{equation}
	\phi(a,y) = \lim_{n\rightarrow\infty} \frac{1}{fn} \log Z_n(a,y) \equiv \lim_{n\rightarrow\infty} \phi_n(a,y) .
\end{equation}

\begin{figure}[t!]
\centering
	\includegraphics[width=0.4\textwidth]{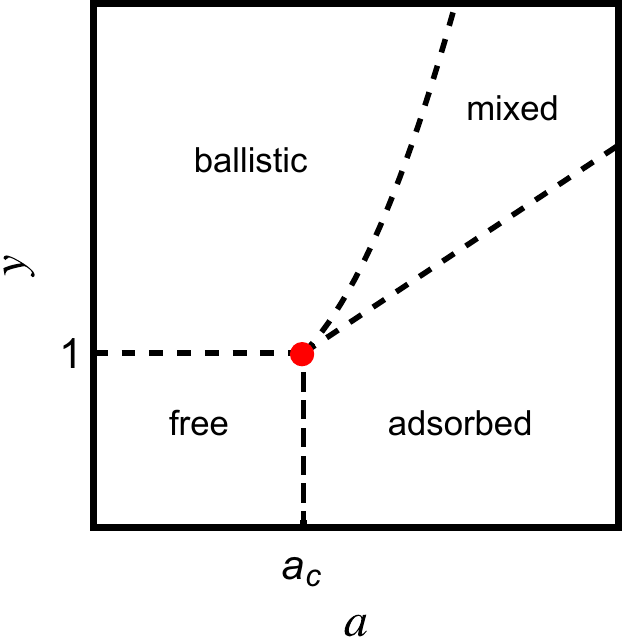}%
	\caption{Expected phase diagram for 3-stars pulled from an adsorbing surface by a force applied at the end of a branch.}%
	\label{fig:PhaseExpected}%
\end{figure}

Recently, Janse van Rensburg and Whittington \cite{Rensburg2018} considered this problem by applying strategy bounds to idealised configuration of $f$-stars. 
They determined that the general free energy in the thermodynamic limit is 
\begin{equation}
	\phi(a,y) = \max\left[\kappa(a),
	\frac{2}{3}\lambda(y) + \frac{1}{3}\log \mu_3,
	\frac{2}{3}\kappa(a) + \frac{1}{3}\lambda(y)\right].
	\label{eq:EndpointStarFreeEnergy}
\end{equation}
where $\kappa(a)$ and $\lambda(y)$ are the free energies of a SAW that is either adsorbed or pulled, but not both, and thus depend only on $a$ and $y$, respectively.
The free energy of a free SAW in $d$ dimensions is $\log\mu_d$, where $\mu_d$ is the connective constant.
Each component of \eref{eq:EndpointStarFreeEnergy} corresponds to one phase.
Respectively these are the fully adsorbed phase, where all three branches are on the surface, the ballistic phase, where two branches are stretched upwards and the third branch is free, and a mixed phase where two branches are adsorbed and the third is pulled away.
Along with a free phase where entropic repulsion dominates both the pulling force and the surface interactions, a schematic phase diagram is illustrated in \fref{fig:PhaseExpected}.
The boundaries between phases are determined from the intersection of the components of \eref{eq:EndpointStarFreeEnergy}.
Together with transitions from the free phase at \smash{$y=1$} and \smash{$a=a_\text{c}$} these boundaries meet at the point \smash{$(a_\text{c},1)$}.

% ===========================================================
% ===========================================================

Recent work on this problem has been mainly theoretical and there is room for numerical work.
Here we seek to verify these theoretical results with Monte Carlo numerical simulations. 
Specifically, we use the flatPERM algorithm \cite{Prellberg2004}, which has proven useful for studying the adsorption of self-avoiding walks and trails \cite{Bradly2018} and, as an athermal simulation, can handle the large range of temperatures needed to map out the phase diagram. 
Although we can calculate the free energy it is more useful to use simulation data to determine thermodynamic averages
\begin{equation}
    \langle Q \rangle(a,y) = \frac{1}{Z_n(a,y)}\sum_{\psi_n \in \scT_n} Q(\psi_n) a^m y^h.
    \label{eq:ThermoQuantity}
\end{equation}
In particular, we are interested in the internal energies
\begin{equation}
    u_n^{(m)} (a,y) = \frac{\avm}{n} \quad\text{and}\quad
	u_n^{(h)} (a,y) = \frac{\avh}{n},
    \label{eq:InternalEnergy}
\end{equation}
which represent the average number of branches that are adsorbed to the surface and the average height of the pulled branch, respectively, and are the natural order parameters for the system. In addition, the relevant size-like quantity for uniform $f$-stars near a surface is the average height of the central node $\avz$.
To locate the phase boundaries we calculate the covariance matrix
\begin{equation} 
	H_n =
	\begin{pmatrix}
	 \frac{\partial^2 \phi_n}{\partial a^2} 	& \frac{\partial^2 \phi_n}{\partial a \partial y}	\\
	 \frac{\partial^2 \phi_n}{\partial y \partial a} 	& \frac{\partial^2 \phi_n}{\partial y^2}
	\end{pmatrix}
	,
	\label{eq:Hessian}
\end{equation}
whose eigenvalues indicate variance in the microcanonical parameters $m$ and $h$.

% ===========================================================
\begin{figure}[t!]
	\includegraphics[width=0.5\textwidth]{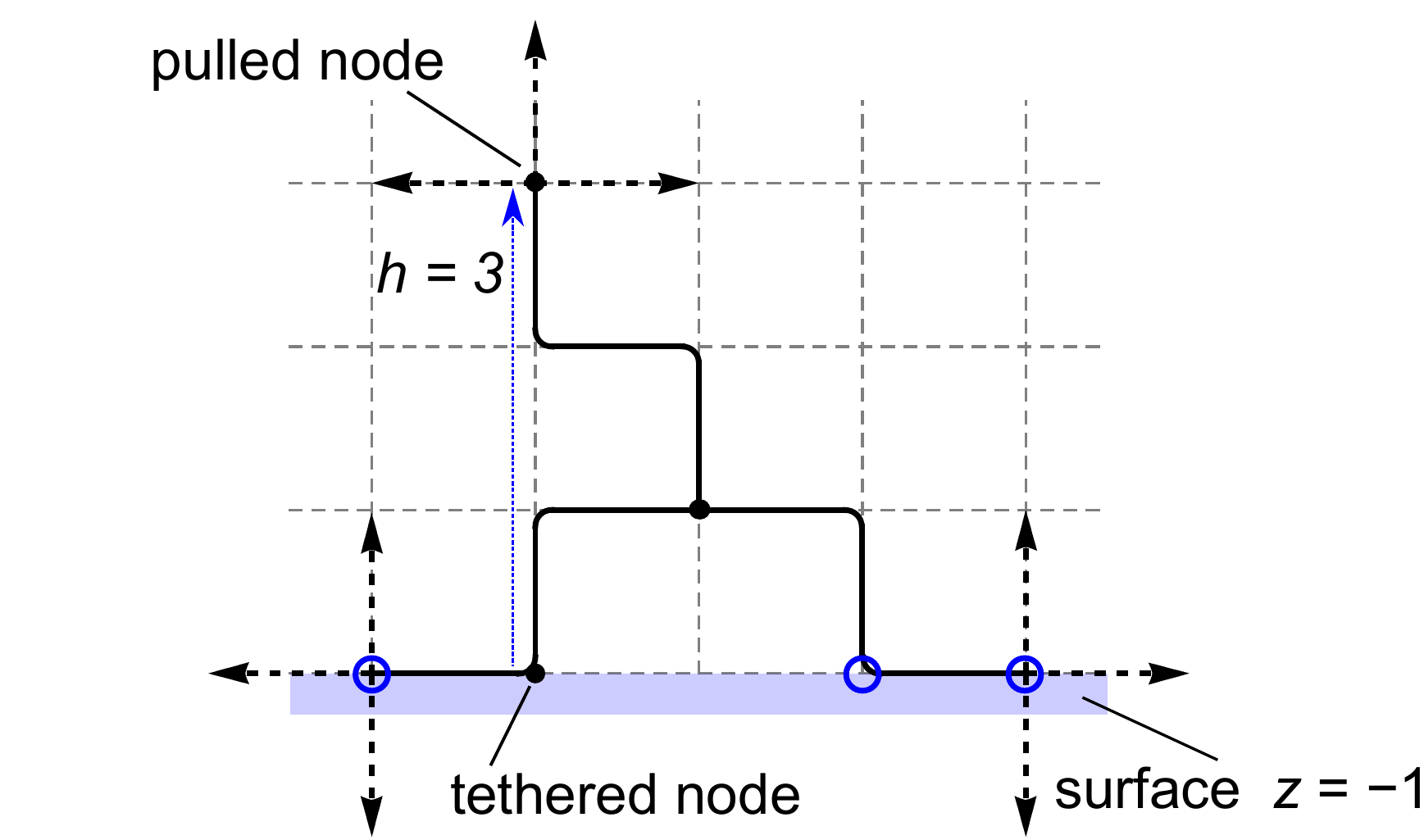}%
	\caption{A 3-star on the square lattice grown to branch length \smash{$n=3$} with height \smash{$h=3$} relative to the surface at \smash{$z=-1$} and \smash{$m=3$} contacts with the surface, excluding the tethered node. 
	Valid growth steps are shown with dashed arrows. 
	Compared to canonical SAWs the only additional restriction is that the pulled branch may not grow through the surface. 
	If another branch grows through the surface, the surface is redefined so that there is always a tether point.
	Blue circles mark monomers that are interacting with the surface. 
	}%
	\label{fig:ValidSteps}%
\end{figure}
% ===========================================================

%%%%%%%%%%%%%%%%%%%%%%%%%%%%%%%%%%%%%%%%%%%%%%%%%%%%%%%%%%%%%
%%%%%%%%%%%%%%%%%%%%%%%%%%%%%%%%%%%%%%%%%%%%%%%%%%%%%%%%%%%%%
\section{Numerical simulation}
\label{sec:Numerical}

The first step in applying a growth algorithm to the problem of simulating $f$-stars is to consider $f$ independent SAWs grown from the same starting point. 
At each step of the growth, the atmosphere of available next steps is calculated for each branch. 
The only difference to SAWs is that calculation of the atmosphere of each branch depends on all other branches.
The atmosphere of the entire $f$-star is the set of all combinations of steps from the atmosphere of each branch.
This prescription would be trivial to implement for models of stars that do not involve a bounding surface but fails when a surface is added and we must therefore alter the algorithm slightly. 

The canonical model for adsorption of lattice polymers begins the walk at the tether point, naturally chosen to be the origin, \smash{${\mathbf r}_0 = (0,0,0)$}. 
The walk is grown from this point with valid occupation sites restricted to the half-space, usually denoted \smash{$x_3 \geq 0$}. 
At each growth step, the atmosphere of available moves can never include a lattice point with \smash{$x_3<0$}. This approach fails for $f$-stars where the central node will not generally lie on the surface, but we also cannot presuppose where the surface will be when the entire star is grown.

\begin{figure}[t!]
\includegraphics[width=\columnwidth]{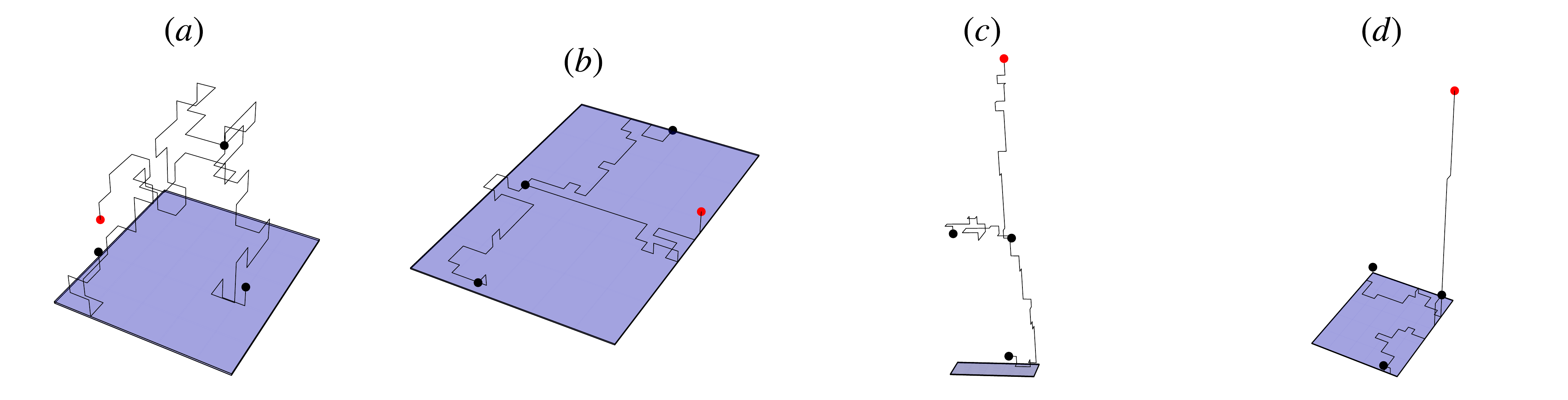}
\caption{Probable configurations of pulled 3-stars with branch length \smash{$n=32$} for (a) the free phase, (b) the adsorbed phase, (c) the ballistic phase and (d) the mixed phase. The pulling force is applied at the red node and the endpoints of the other branches are not in the surface so the star is tethered at some other point.}%
\label{fig:BestConfigurations}%
\end{figure}

The solution we have employed is to keep the central node at \smash{${\mathbf r}_0 = (0,0,0)$} and alter the growth steps slightly, as visualised in \fref{fig:ValidSteps}.
Initially, the surface is set to \smash{$x_3=0$} and one of the branches is tagged as the pulled branch.
As with adsorbing SAWs, nodes on the other side of the surface are excluded from the atmosphere of the pulled branch.
However, the other $f-1$ branches do not have this restriction and may grow through the boundary at any step.
If a branch grows to the other side of the surface then the surface is redefined so that the entire star is in one half-space.
This is equivalent to a translation of the $f$-star, which preserves its properties.
Note that this is an extension to the set of moves for growing configurations and says nothing about whether the resulting configuration will be accepted as valid by the Monte Carlo algorithm (e.g.~whether the configuration will be pruned or enriched according to flatPERM; see below). 
With this prescription we have an ergodic algorithm for growing $f$-stars tethered to a surface.
The disadvantage is that we are no longer modeling $f$-stars that are tethered to the surface at the endpoint of one of the branches.
Instead, a branch that grows through the surface may at a later step grow away from the surface, leaving the tether point at an arbitrary interior point in the branch.
Figure \ref{fig:BestConfigurations} shows probable, i.e.~highly weighted, configurations for each of the four expected phases produced by the simulation at branch length \smash{$n=32$}. 
Each example has endpoints that are not in the surface, thus the tether point is an arbitrary interior node. 
In the next section we present a more detailed argument for using this model.

We use the flatPERM algorithm \cite{Prellberg2004}, an extension of the pruned and enriched Rosenbluth method (PERM) \cite{Grassberger1997}. 
In the standard case of a SAW the simulation works by growing a walk on a given lattice up to some maximum length. 
At each step the cumulative Rosenbluth \& Rosenbluth weight \cite{Rosenbluth1955} of the walk is compared with the current estimate of the density of states $W_{nmh}$. 
If the current state has relatively low weight (e.g.~by being trapped or reaching the maximum length) the walk is `pruned' back to an earlier state. 
On the other hand, if the current state has relatively high weight, then microcanonical quantities are measured and $W_{nmh}$ is updated. 
The state is then `enriched' by branching the simulation into several possible further paths (which are explored when the current path is eventually pruned back). 
When all branchings are pruned a new iteration is started from the origin.

% ===========================================================
% ===========================================================

FlatPERM enhances this method by altering the prune or enrich choice such that the sample histogram is flat in the microcanonical parameters $n$, $m$ and $h$. 
As well as adapting the atmosphere of available growth steps, there are some necessary alterations to the flatPERM algorithm in the case of $f$-stars. 
The prune and enrich cycle is applied to the $f$-star as a whole, not to individual branches. 
This means that all branches have equal length at all stages of the growth and if a single branch is trapped, the whole $f$-star is trapped and all branches are pruned back to a shorter length. 

The main output of the simulation is the density of states $W_{nmh}$ of $f$-stars with branch length $n$ with $m$ contacts with the surface and pulled at height $h$, for all $n$ up to some fixed maximum $n_\text{max}$. 
Thermodynamic quantities are then given by the weighted sum
\begin{equation}
	\langle Q \rangle_n(a,y) = \frac{\sum_{m,h} Q(n,m,h) W_{nmh} a^m y^h }{\sum_{m,h} W_{nmh} a^m y^h}.
    \label{eq:DoSQuantity}
\end{equation}
The probability distribution of the number of contacts at a given temperature and interaction strength (i.e.~given $a$ and $y$) is 
\begin{equation}
	P(m) = \frac{\sum_{h} W_{nmh} a^m y^h }{\sum_{m,h} W_{nmh} a^m y^h}.
    \label{eq:DoSProbability}
\end{equation}
The probability distribution for the height of the pulled node, $P(h)$ is defined similarly.
Size-like microcanonical quantities such as $\avz$ are also calculated during the simulation similar to the weights $W_{nmh}$.

%%%%%%%%%%%%%%%%%%%%%%%%%%%%%%%%%%%%%%%%%%%%%%%%%%%%%%%%%%%%%
%%%%%%%%%%%%%%%%%%%%%%%%%%%%%%%%%%%%%%%%%%%%%%%%%%%%%%%%%%%%%
\section{3-stars tethered at an arbitrary interior point}
\label{sec:FreeEnergy}

Before looking at results of the simulations, we must consider whether the model of $3$-stars that are tethered to a surface at an arbitrary interior node is the same or similar to the case where the star is tethered at an endpoint.
Intuitively, the location of the tether is a minor point.
For stars in the fully adsorbed or mixed phases the interaction with the surface is strong, a large fraction of the nodes in the non-pulled branches are in contact with the surface and it is likely that the endpoint is one of the nodes on the surface.
Thus the same configuration is valid in both the endpoint-tethered and interior-tethered models with otherwise identical properties.
For stars in the ballistic phase, as the pulling force is increased the tether point will preferentially be towards the endpoint so the free energy of the two models should converge at least in the limit of large force.
In the free phase, the surface has no effect in the thermodynamic limit so the location of the tether is unimportant.

In this section we follow the proof of the free energy, \eref{eq:EndpointStarFreeEnergy}, given in \cite{Rensburg2018}, which assumes the $f$-star is tethered at the endpoint of one of the branches.
There are only a few minor alterations for the case of arbitrary interior-tethered $f$-stars, and thus we omit some of the technical details.
The following discussion is specific to 3-stars, but is easily extended to \smash{$3 \leq f \leq z$}, where $z$ is the coordination number of the lattice, for example the three dimensional simple cubic lattice has \smash{$z=6$}.

We begin with notation for SAWs since they are the constituent components of 3-stars.
Let $\lambda(y)$ be the free energy of a SAW tethered to a surface at one end and pulled at the other with force $F$ (and conjugate Boltzmann weight $y$), but without interactions with the surface.
There is a free phase where \smash{$\lambda(y)=\log\mu_3$} and a continuous transition at \smash{$y=1$} to a ballistic phase, which has asymptotic behaviour \smash{$\lambda(y)\sim\log(y)$} for large $y$ \cite{Rensburg2013}.
Next, let $\kappa(a)$ be the free energy of a SAW and subject to attractive interactions with the surface.
In the free phase \smash{$\kappa(a)=\log\mu_3$} and there is a continuous transition at some \smash{$a_\text{c}>1$} to the adsorbed phase, which has asymptotic behaviour \smash{$\kappa(a)\sim\log(a)+\log\mu_2$} for large $a$ \cite{Rychlewski2011}.
The free energy of a SAW with neither pulling nor adsorption is $\log\mu_3$.

Turning to $3$-stars, the first part is to consider regimes of different $a$ and $y$, where there is either pulling or adsorption, or both.
In all cases, a lower bound to the free energy of 3-stars is straightforward.
In our model we include all 3-stars with one branch tethered to the surface at an arbitrary interior point.
The partition function for this model, \eref{eq:Partition}, includes terms for configurations that are tethered at the endpoint of a branch and configurations that are not.
Since $a$ and $y$ are non-negative, the sum of endpoint-tethered terms are less than the entire partition function.
In the limit of large $n$, the free energy of endpoint-tethered 3-stars, \eref{eq:EndpointStarFreeEnergy}, is therefore a lower bound on the free energy of 3-stars tethered at an arbitrary interior point.

% ===========================================================
\begin{figure}[t!]
\includegraphics[width=\columnwidth]{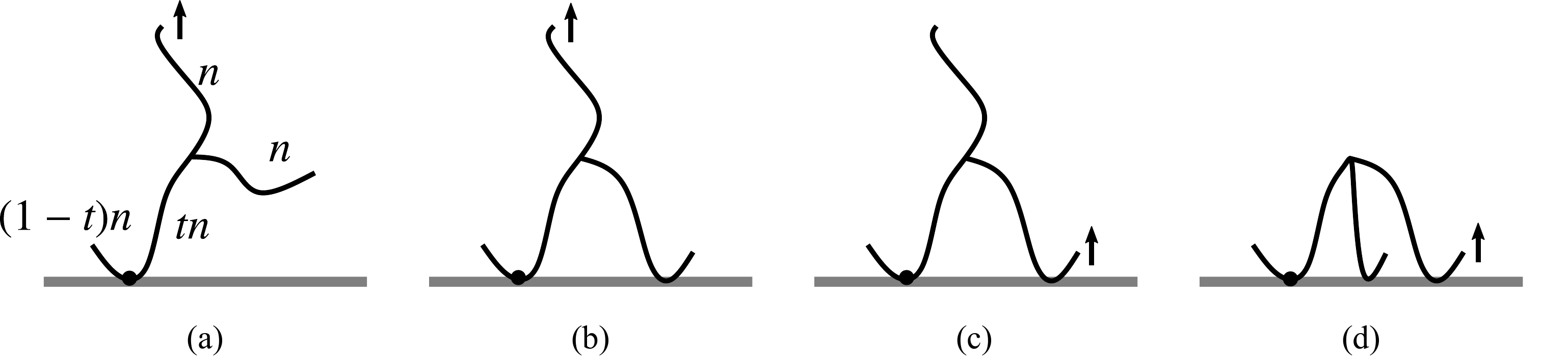}
\caption{Idealised configurations of pulled 3-stars with one branch tethered at an arbitrary interior node and another branch being pulled at the endpoint.}%
\label{fig:ConfigurationCases}%
\end{figure}
% ===========================================================

%%%%%%%%%%%%%%%%%%%%%%%%%%%%%%%%%%%%%%%%%%%%%%%%%%%%%%%%%%%%%
\subsection{Bounds for $a>1$, $y>1$}

For the general regime where there is both pulling and adsorption, \smash{$a>1$} and \smash{$y>1$}, the free energy will be composed of both $\kappa(a)$ and $\lambda(y)$.
The free energy of 3-stars tethered at the endpoint of a branch, \eref{eq:EndpointStarFreeEnergy} is a lower bound on the free energy of 3-stars tethered at an arbitrary interior point:
\begin{equation}
	\liminf_{n\to\infty}\frac{1}{3n}\log Z_n(a,y) \geq \max\left[\kappa(a),
	\frac{2}{3}\lambda(y) + \frac{1}{3}\log \mu_3,
	\frac{2}{3}\kappa(a) + \frac{1}{3}\lambda(y)\right].
	\label{eq:LowerBoundAll}
\end{equation}

Upper bounds are found by treating the branches as independent SAWs, each contributing a term to the free energy of the 3-star.
We look at the idealised configurations shown in \fref{fig:ConfigurationCases}.
The configurations are distinguished by the number of branches in contact with the surface and which branch is being pulled.

For configuration (a) the 3-star can be decomposed into a free branch, a pulled branch with no points in the surface, and a pulled branch with points in the surface and tethered at an arbitrary interior point.
The latter can be further decomposed into an unpulled SAW of length $(1-t)n$ and a pulled SAW of length $tn$, both tethered at an endpoint and possibly having contacts with the surface.
The unpulled dangling tail does not feel the pulling force so it has free energy $\kappa(a)$.
The portion between the tether point and the central node, which transmits the pulling force, has free energy that is a linear combination of $\lambda(y)$ and $\kappa(a)$, and is therefore bounded by $\max\left[ \kappa(a), \lambda(y) \right]$.
The free energy of the tethered branch is the sum of these two terms, weighted by their relative length: \smash{$(1-t)\kappa(a) + t\max\left[ \kappa(a), \lambda(y) \right]$}.
Considered as a component of the 3-star, $t$ is not an {\em a priori} parameter, and the 3-star partition function includes configurations for all values of $t$.
Thus in order to find an upper bound we are now able to optimise with respect to $t$. 
This means that the free energy of the tethered branch is bounded by the maximum of the two terms, and since one already includes the other, we can say that the free energy of the tethered branch is bounded by $\max\left[ \kappa(a), \lambda(y) \right]$.
When we add the contribution from the pulled branch and the free branch, the total upper bound to the 3-stars free energy based on configuration (a) is therefore
\begin{equation}
	\limsup_{n\to\infty}\frac{1}{3n}\log Z_n(a,y) \leq \frac{1}{3}\left[
	\max\left[\kappa(a),\lambda(y)\right] + \lambda(y) + \log\mu_3\right].
	\label{eq:UpperBoundA}
\end{equation}

For configuration (b) decompose the 3-star into the pulled branch and the two branches contacting the surface.
The two branches in the surface are decomposed not with respect to the central node, but via the tether point, and form a loop of length $(1+t)n$ tethered at one of its endpoints, which is being pulled, and an unpulled tail of length $(1-t)n$, also tethered at its endpoint.
The free energy of loops pulled at their highest point is $\lambda(\sqrt{y})$ \cite{Guttmann2018}.
In this case, the loop is being pulled at the central node of the star, which may not be the highest point of the loop, so $\lambda(\sqrt{y})$ is an upper bound.
To see this, consider that each pulled loop is weighted by $y^h$ in the partition function, where $h$ is the height of the node where the pulling force is applied.
For \smash{$y>1$} all such weights are bounded by $y^s$, where $s$ is the highest point of the loop, so $\lambda(\sqrt{y})$ is a bound on the partition function of our loop.
The loop also feels the surface interaction so its free energy is a linear combination of $\kappa(a)$ and $\lambda(\sqrt{y})$, which is bounded by $\max\left[ \kappa(a),\lambda(\sqrt{y}) \right]$.
The tail does not feel the pulling force and is in contact with the surface so it has free energy $\kappa(a)$.
Together, the loop and tail has free energy bounded by $(1+t)\max\left[ \kappa(a), \lambda(\sqrt{y}) \right]+ (1-t)\kappa(a)$.
As before we can optimise over the parameter $t$.
The maximum is either \smash{$\max\left[ \kappa(a), \lambda(\sqrt{y}) \right]$} or $\kappa(a)$, but the latter is redundant since it appears in the former.
Hence the loop and tail component contributes $2\max\left[ \kappa(a), \lambda(\sqrt{y}) \right]$ to the bound on the free energy of the star.
Along with the pulled branch which is not in contact with the surface, thus contributing $\lambda(y)$, we obtain the following upper bound for the 3-star in configuration (b)
\begin{equation}
	\limsup_{n\to\infty}\frac{1}{3n}\log Z_n(a,y) \leq \frac{1}{3}\left[
	2\max\left[ \kappa(a), \lambda(\sqrt{y}) \right] + \lambda(y)\right].
	\label{eq:UpperBoundB}
\end{equation}

Case (c) is similar to case (b) except now the force is applied to a branch that is in contact with the surface, while the third branch is free.
The pulled branch depends on $a$ and $y$ and contributes $\max\left[ \kappa(a),\lambda(y) \right]$
The tethered branch is not subject to a force so its free energy contribution is $\kappa(a)$, independent of the location of the tether.
The upper bound associated with this configuration is therefore
\begin{equation}
	\limsup_{n\to\infty}\frac{1}{3n}\log Z_n(a,y) \leq \frac{1}{3}\left[
	\max\left[ \kappa(a),\lambda(y) \right] + \kappa(a) + \log\mu_3\right].
	\label{eq:UpperBoundC}
\end{equation}

%\item 
Lastly, consider configuration (d) where all three branches are in contact with the surface.
Two branches only interact with the surface so contribute free energy $\kappa(a)$ while the contribution of the pulled branch may also depend on $y$ so we have the bound
\begin{equation}
	\limsup_{n\to\infty}\frac{1}{3n}\log Z_n(a,y) \leq \frac{1}{3}\left(
	2\kappa(a) + \max\left[ \kappa(a),\lambda(y) \right] \right).
	\label{eq:UpperBoundD}
\end{equation}
%\end{enumerate}

For a total upper bound, we note that \smash{$\lambda(\sqrt{y})\leq\tfrac{1}{2}\left[\lambda(y)+\log\mu_3\right]$} since $\lambda(y)$ is a convex function of $\log y$.
We also know that $\kappa(a)\geq\log\mu_3$, for \smash{$a>1$}, and $\lambda(y)\geq\log\mu_3$, for \smash{$y>1$}.
We can now combine Eqs.~\eqref{eq:UpperBoundA}, \eqref{eq:UpperBoundB}
, \eqref{eq:UpperBoundC} 
and \eqref{eq:UpperBoundD} to obtain
\begin{equation}
	\limsup_{n\to\infty}\frac{1}{3n}\log Z_n \leq \max\left[\kappa(a),
	\frac{2}{3}\lambda(y) + \frac{1}{3}\log \mu_3,
	\frac{2}{3}\kappa(a) + \frac{1}{3}\lambda(y)\right].
	\label{eq:UpperBoundAll}
\end{equation}

This is precisely the same as the lower bound, \eref{eq:LowerBoundAll}, and therefore the free energy of 3-stars with an arbitrary interior tether is the same as the free energy of 3-stars with an endpoint tether for \smash{$a>1$} and \smash{$y>1$}:
\begin{equation}
	\phi(a,y) = \max\left[\kappa(a),
	\frac{2}{3}\lambda(y) + \frac{1}{3}\log \mu_3,
	\frac{2}{3}\kappa(a) + \frac{1}{3}\lambda(y)\right].
	\label{eq:InteriorStarFreeEnergy}
\end{equation}

%%%%%%%%%%%%%%%%%%%%%%%%%%%%%%%%%%%%%%%%%%%%%%%%%%%%%%%%%%%%%
\subsection{Bounds for $a=1$ or $y=1$}
\label{sec:SingleParameterCases}

We briefly consider the cases where there is pulling or adsorption, but not both.
In either case, because $\kappa(a)$ and $\lambda(y)$ are convex functions, the result of the previous section for \smash{$a>1$} and \smash{$y>1$} serves as an upper bound.
Further, since \smash{$\kappa(a)\geq\log\mu_3$} and \smash{$\lambda(y)\geq\log\mu_3$}, we can set either \smash{$a=1$} or \smash{$y=1$} in \eref{eq:InteriorStarFreeEnergy} to obtain
\begin{subequations}\label{eq:SingleParameterUpperBounds}
\begin{align}
	\limsup_{n\to\infty}\frac{1}{3n}\log Z_n(1,y) &\leq \frac{2}{3}\lambda(y) + \frac{1}{3}\log\mu_3,	\\
	\limsup_{n\to\infty}\frac{1}{3n}\log Z_n(a,1) &\leq \kappa(a).
\end{align}
\end{subequations}
As with the general case, the free energy of endpoint-tethered 3-stars is a lower bound on the free energy of 3-stars tethered at an arbitrary interior point.
The results of \cite{Rensburg2018} are the same as \eref{eq:SingleParameterUpperBounds} and so the free energy is the same for \smash{$a=1$} or \smash{$y=1$}.
Hence, the free energy is indifferent to the location of the arbitrary tether point and our model may be used to simulate the problem of $f$-stars tethered to a surface.

%%%%%%%%%%%%%%%%%%%%%%%%%%%%%%%%%%%%%%%%%%%%%%%%%%%%%%%%%%%%%
%%%%%%%%%%%%%%%%%%%%%%%%%%%%%%%%%%%%%%%%%%%%%%%%%%%%%%%%%%%%%
\section{Simulation Results}
\label{sec:Results}

% ===========================================================
\begin{figure*}[t!]
\centering
	\begin{tabular}{ccc}
	\includegraphics[width=0.33\linewidth]{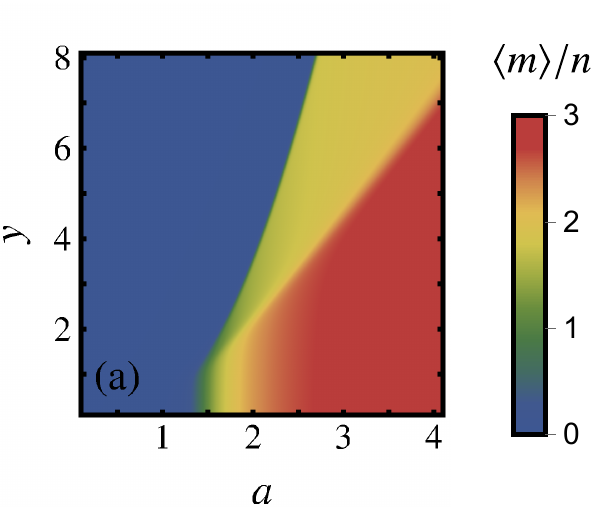}
	\label{fig:PhaseAdsorbed}
	&
	\includegraphics[width=0.33\linewidth]{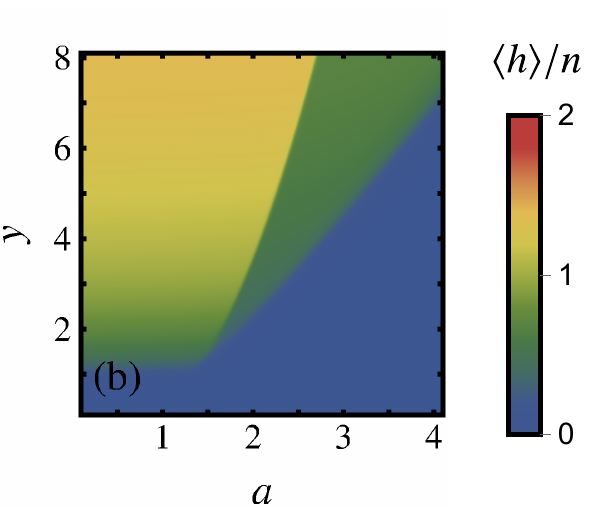}
	\label{fig:PhasePulled}
	&
	\includegraphics[width=0.33\linewidth]{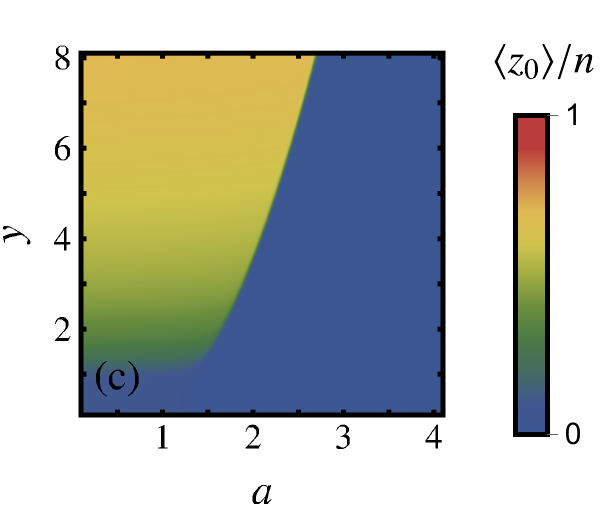}
	\label{fig:PhaseCentral}
	\end{tabular}
	\caption{The internal energies (a) \smash{$\avm/n$} and (b) \smash{$\avh/n$}, and (c) height of the central node \smash{$\avz/n$}, for branch length \smash{$n=128$}. 
	Four phases are apparent: the free phase for \smash{$a\leq a_\text{c}$} and \smash{$y\leq 1$}; an adsorbed phase at high $a$ and small $y$; a ballistic phase at high $y$ and low $a$; and a mixed phase between the adsorbed and ballistic phases.
	}
	\label{fig:PhaseDiagram}
\end{figure*}
% ===========================================================

In this work we have simulated pulled 3-stars with branch lengths up to \smash{$n \leq 128$} using flatPERM as described above.
We ran ten independent simulations with \smash{$10^4$} iterations each and averaged the results, obtaining a total of $5.3\times10^{10}$ samples at maximum branch length \smash{$n=128$}.
To map out the phase diagram we look at the internal energies, \eref{eq:InternalEnergy}. 
Figure \ref{fig:PhaseDiagram} shows (a) the average number of adsorbed branches \smash{$\avm/n$}, (b) the average height of the pulled branch \smash{$\avh/n$}, and (c) the average height of the central node \smash{$\avz/n$},  for branch length \smash{$n=128$}. 
Collectively, these quantities show the four phases: free, adsorbed, ballistic and mixed.

\subsection{Phase diagram}
\label{sec:PhaseDiagram}

The free phase is bounded by the adsorption transition point \smash{$a\leq a_\text{c}$} and \smash{$y\leq 1$}, where the force changes from a pull away from the surface into a local push towards the surface. Note that this matches the known result for SAWs \cite{Beaton2015}.
Within this phase both the expected number of surface contacts and the average height of the pulled (or pushed in this case) node is zero.
The tethering of one branch is the only restriction on the configuration of the 3-star polymer.
Seeing $\avm$ and $\avh$ vanish also indicates that the free energy is expected to be independent of $a$ and $y$.

As $a$ is increased the system undergoes a transition to the adsorbed phase at a critical temperature \smash{$a_\text{c}>1$}.
Beyond the critical point, the average number of surface contacts $\avm$ quickly approaches the maximum $3 n$ while $\avh$ and $\avz$ are suppressed to zero.
This indicates that all three branches are adsorbed to the surface.
Further, within the adsorbed phase the free energy is independent of $y$, although, as $a$ increases, the maximum value of $y$ for which this is true increases from \smash{$y=1$} in the free phase.

If we again start in the free phase and increase $y$ the system enters the ballistic phase at \smash{$y=1$}, where the thermodynamics depends only on the single constant pulling force.
This phase is characterised by $\avm$ tending to zero while both $\avh$ and $\avz$ are of order $n$.
The expected configuration is that the pulled and tethered branches are stretched out perpendicular to the surface while the third branch assumes a disordered coil configuration relative to the central node.
We note that even in this phase the $\avh$ only slowly approaches its maximum value $2 n$ as $y$ is increased, whereas $\avz$ more quickly finds its maximum value $n$.
Analogously to the adsorbed phase, now the free energy is independent of $a$.

% ===========================================================
\begin{figure*}[t!]
\centering
	\includegraphics[width=0.4\linewidth]{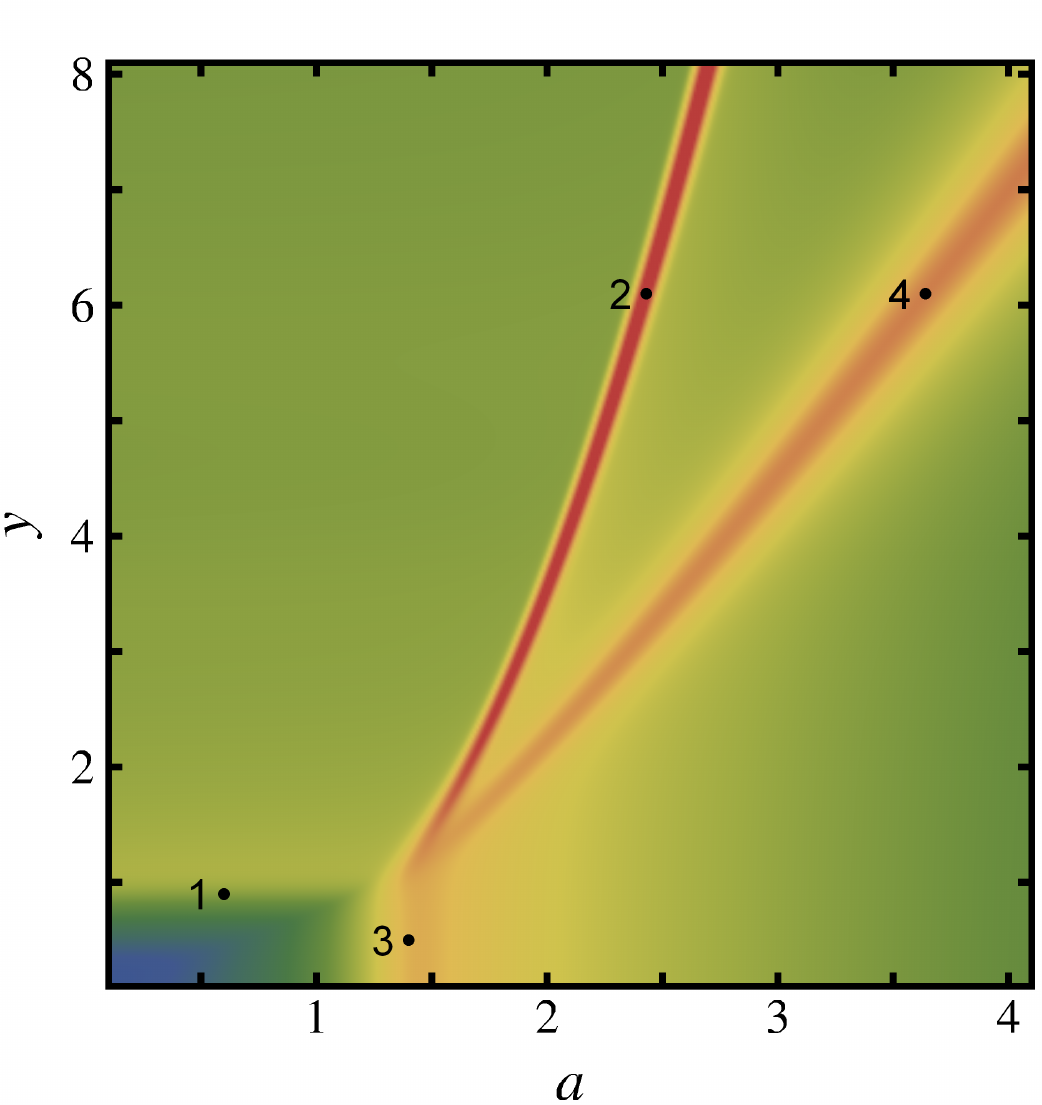}
	\vspace{0.3cm}
	\includegraphics[width=\linewidth]{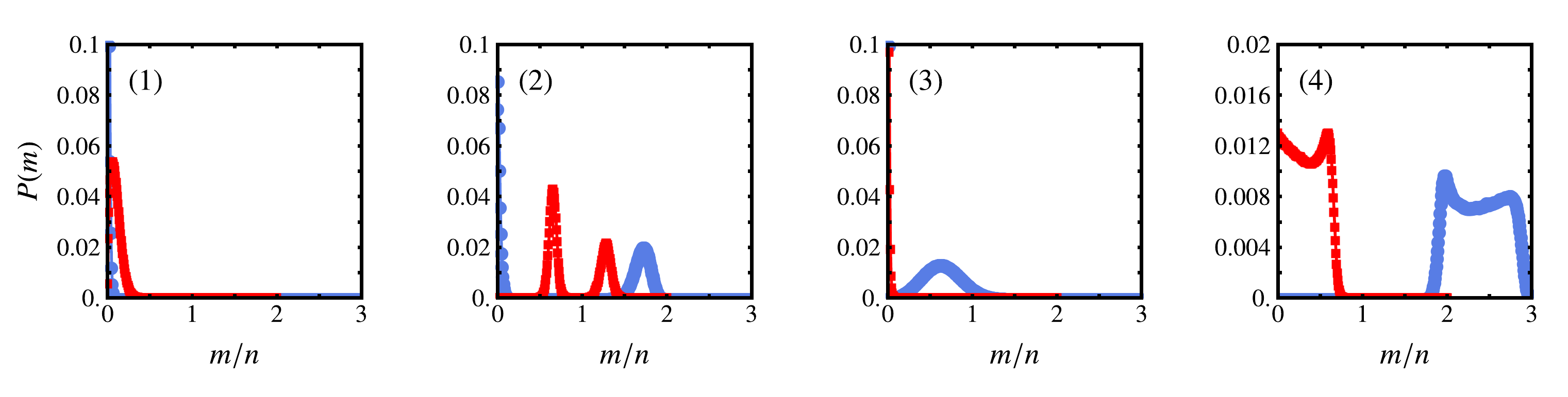}
	\caption{(Top) Density plot of the logarithm of the largest eigenvalue of the Hessian matrix of the free energy for \smash{$n=128$}. 
	The phase boundaries are clearly visible as lines of high variance. 
	(Bottom) The distribution of number of adsorbed branches (blue) and the distribution of height of pulled node (red) at points of interest as marked on the density plot. 
	Note the maximum possible values are \smash{$m/n=3$} and \smash{$h/n=2$}.
	}

	\label{fig:HessianAndDistributions}
\end{figure*}
% ===========================================================

Between the adsorbed and ballistic phases is a mixed phase where $\avm$ is of order $2n$ and $\avh$ is of order $n$. 
At the same time we see that \smash{$\avz$} vanishes.
The free energy in this phase thus depends on both $a$ and $y$.
This indicates that one branch has been pulled away from the surface but two remain adsorbed and in particular the central node is still on the surface.
The appearance of this mixed phase is what makes lattice stars distinct from linear SAWs pulled from an adsorbing surface, but it is also seen in self-avoiding polygon models of ring polymers \cite{Beaton2017}. 

We now turn to the nature of the phase transitions.
The discussion above indicates that the likely configurations in the free, adsorbed, mixed and ballistic phases involve respectively, zero, three, two and zero branches adsorbed to the surface.
We expect to see that that the transitions to the mixed phase are first order.
In \fref{fig:HessianAndDistributions} we show a density plot of the logarithm of the largest eigenvalue of $H_n$ using data for \smash{$n=128$}. 
The adsorbed-mixed and ballistic-mixed phase boundaries are distinctly visible indicating fairly sharp transitions.

% ===========================================================
\begin{figure*}[t!]	
\begin{tabular}[t]{cccc}
	\includegraphics[width=0.33\linewidth]{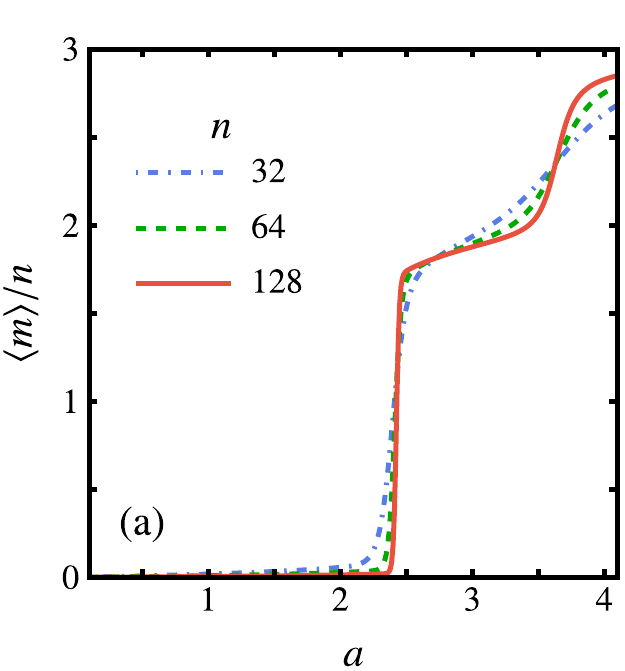}&&&
	\includegraphics[width=0.33\linewidth]{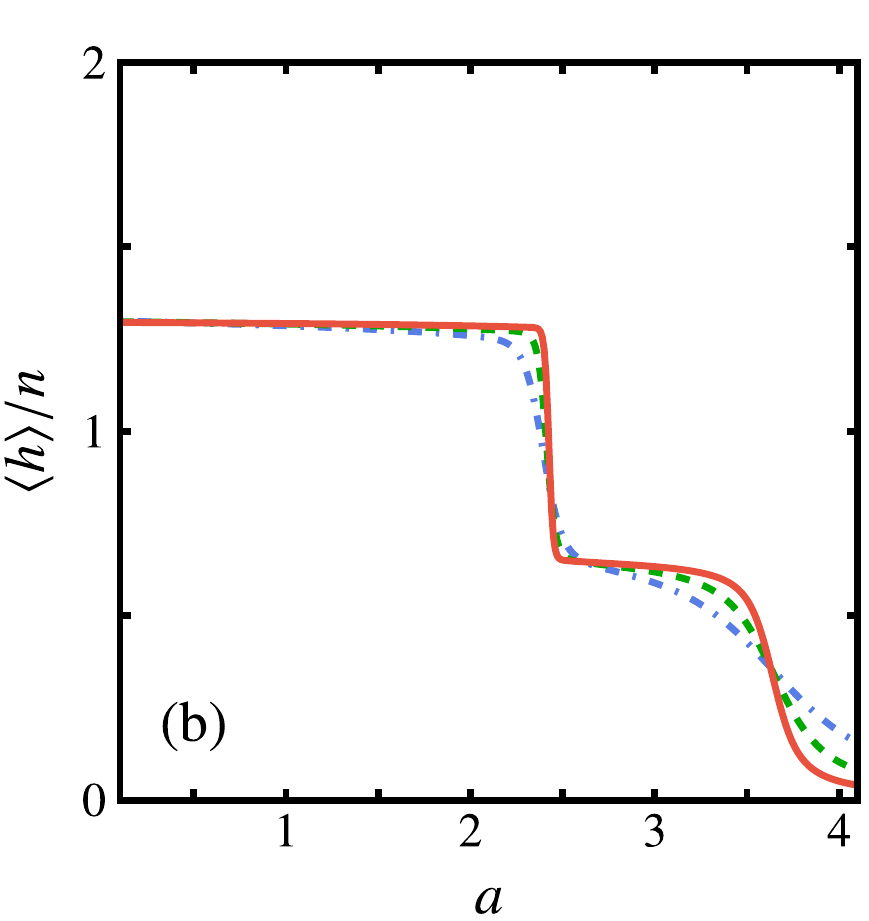}
\end{tabular}
\begin{tabular}[t]{cccc}
	\includegraphics[width=0.33\linewidth]{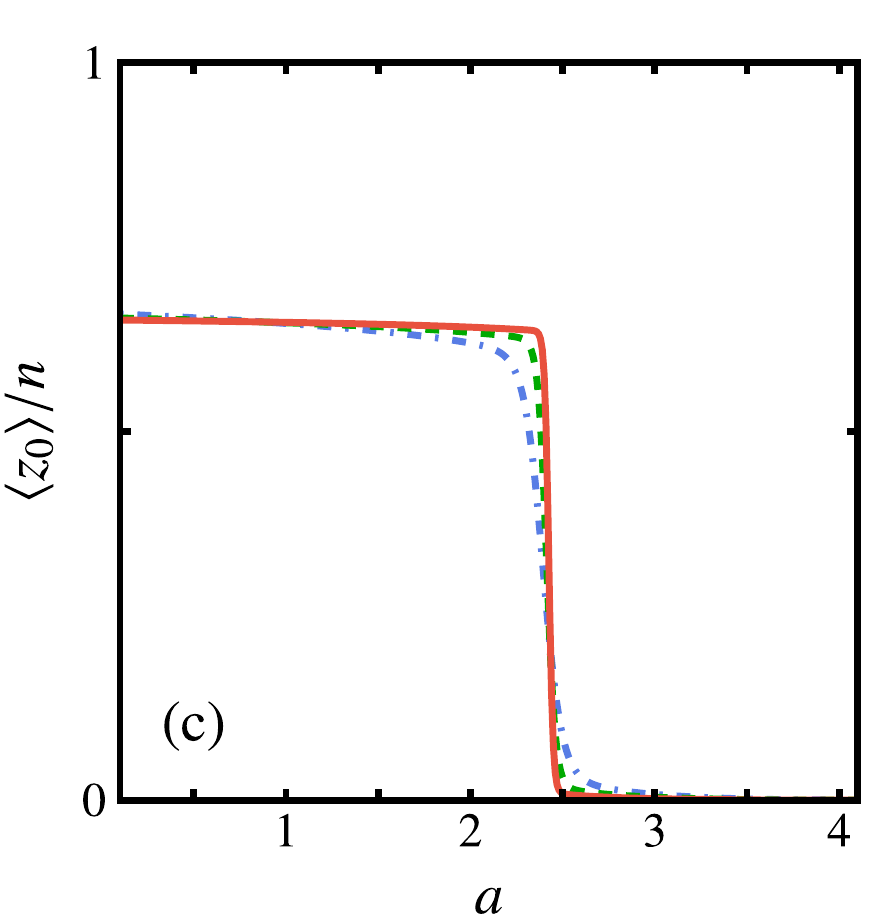}&&&
	\includegraphics[width=0.33\linewidth]{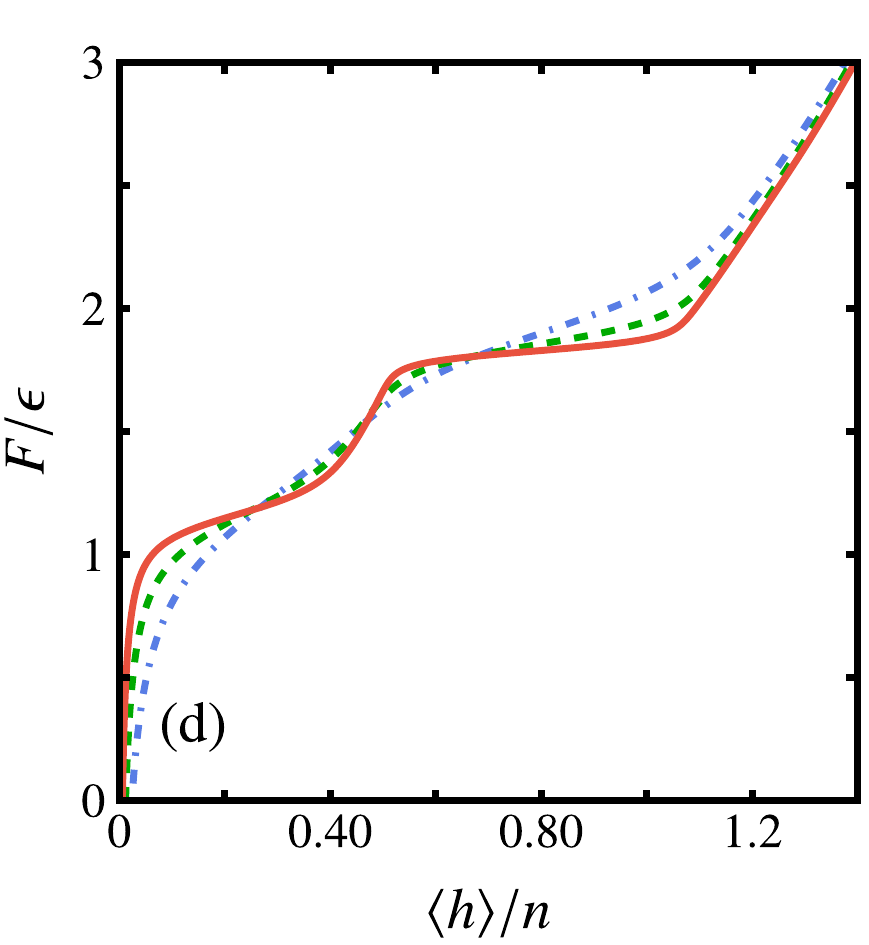}
\end{tabular}

	\caption{Internal energies (a) $\avm/n$ and (b) $\avh/n$ and (c) height of central node $\avz/n$ as a function of $a$ for fixed $y=6.1$ and \smash{$n=32,64,128$}. 
	The ballistic-mixed transition is visible at \smash{$a\approx 2.4$} in all three quantities while and the weaker adsorbed-mixed transition at \smash{$a\approx 3.5$} does not alter the height of the central node.
	(d) Force-extension graph at fixed temperature corresponding to fixed \smash{$a=2$}.
	}

	\label{fig:EnergySlices}
\end{figure*}
% ===========================================================

The type of transition can be determined by looking at the underlying distributions $P(m)$ and $P(h)$ of the number of surface contacts and the height of the pulled node, respectively. 
These distributions are calculated directly from the density of states $W_{nmh}$ output by the simulation at fixed $a$ and $y$ as marked on the density plot. 
The distributions at several points of interest in the $a$-$y$ plane are shown in \fref{fig:HessianAndDistributions} and are indicative of all points along the phase boundaries.
We see that for the adsorbed-mixed and ballistic-mixed phase boundaries the distributions $P(m)$ and $P(h)$ are bimodal, indicating that both transitions are first order. 
In contrast, the distributions near the free-ballistic and free-adsorbed boundaries are not bimodal and these transitions are continuous as expected from the case of SAWs.

To confirm that the ballistic-mixed and adsorbed-mixed transitions are first order we plot in \fref{fig:EnergySlices} the internal energies (a) $\avm/n$ and (b) $\avh/n$ as well as (c) the height of the central node $\avz/n$ as a function of $a$ for fixed \smash{$y=6.1$} at several values of $n$.
This is a horizontal slice through the phase diagram corresponding to points 2 and 4 in \fref{fig:HessianAndDistributions}.
As $n$ increases, the ballistic-mixed transition (\smash{$a\approx 2.4$}) displays a sharply defined latent heat.
The adsorbed-mixed transition (\smash{$a\approx 3.5$}) is broader but it is clear that the trend for larger $n$ is also the emergence of a latent heat.
The ballistic-mixed transition is also visible in the height of the central node $\avz/n$ but the adsorbed-mixed transition is not, as expected

Another viewpoint is to consider that atomic-force microscopy experiments are performed at a fixed temperature $T$ below the adsorption transition temperature while measuring the force $F$ and extension $h$ \cite{Alvarez2009}.
In our parameterization this corresponds to a plot of $F$ (in units of $\epsilon$) versus the average height of the pulled node $\avh/n$, where \smash{$F/\epsilon=\log y/\log a$}.
In \fref{fig:EnergySlices}(c) we show a force-extension plot at fixed temperature corresponding to \smash{$a=2$}.
As $n$ is increased there is a clear emergence of plateaus in force $F$ as the extension is increased.
These results indicate that the ballistic-mixed and adsorbed-mixed transitions are first-order.

\subsection{Phase boundaries}
\label{sec:PhaseBoundaries}

\begin{figure}[t!]
	\centering
	\includegraphics[width=0.5\columnwidth]{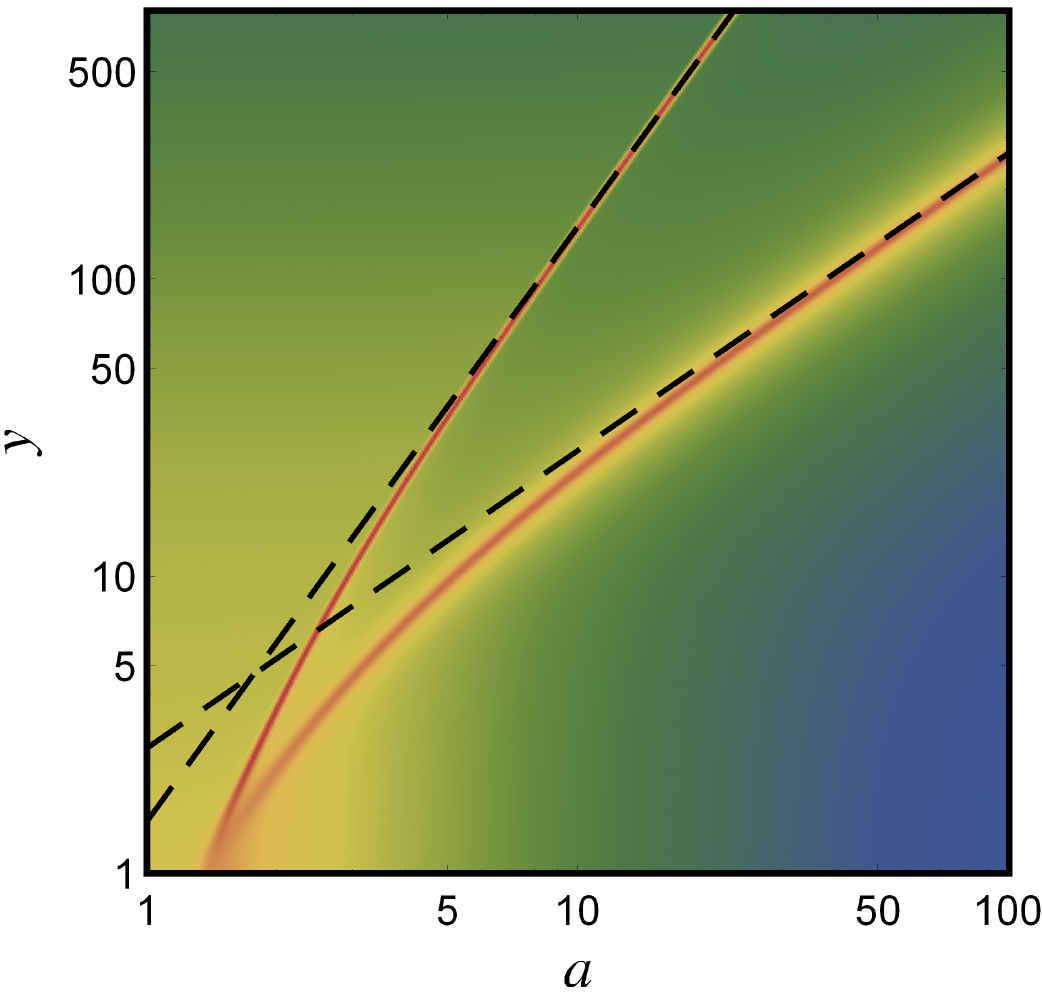}
	\caption{The logarithm of largest eigenvalue of the Hessian covariance matrix for large $a$ and $y$, shown on a logarithmic scale. Dashed lines indicate the expected asymptotic behaviors from \eref{eq:Asymptotes}.}%
	\label{fig:PhaseAsymptotes}%
\end{figure}

The position of the phase boundaries are determined by the intersection of the terms in \eref{eq:EndpointStarFreeEnergy}, that is, the solutions of 
\begin{subequations}\label{eq:BoundaryEquations}
\begin{align}
	&\lambda(y) + \log \mu_3 = 2\kappa(a) 	&&\text{(ballistic-mixed)}	\\
	&\kappa(a) = \lambda(y)  				&&\text{(adsorbed-mixed)}.
\end{align}
\end{subequations}
We do not have accurate knowledge of $\kappa(a)$ and $\lambda(y)$ for all $a$ and $y$, but we can use the asymptotic behavior (see \sref{sec:FreeEnergy}) to obtain the phase boundaries for large $a$ and $y$:
\begin{subequations}\label{eq:Asymptotes}
\begin{align}
	y &\sim \frac{\mu_2^2}{\mu_3} a^2	&&\text{(ballistic-mixed)}	\\
	y &\sim \mu_2 a 					&&\text{(adsorbed-mixed)}.
\end{align}
\end{subequations}
In \fref{fig:PhaseAsymptotes} we show the logarithm of the largest eigenvalue of $H_n$ for \smash{$n=128$} on a log-log plot for a larger range of $a$ and $y$. 
The asymptotic forms of \eref{eq:Asymptotes} are superimposed as dashed lines with slope 1 and 2 using known values for $\mu_2$ and $\mu_3$ \cite{Jensen1998,Clisby2013}.
It is immediately clear that both the ballistic-mixed and adsorbed-mixed boundaries follow the expected forms for large $a$ and $y$ as expected from \cite{Rensburg2018}.
At smaller $a$ and $y$ both boundaries approach the asymptote from below, in accordance with the fact that the free energies $\kappa(a)$ and $\lambda(y)$ are convex functions.
The deviation also indicates the re-entrance of the mixed phase at low temperature for fixed pulling force.
However, further study of the position of the boundaries at smaller $a$ and $y$ requires highly accurate knowledge of $\kappa(a)$ and $\lambda(y)$, and is beyond the scope of this study.
In this regime it is possible that the location of the tether at an arbitrary interior point, rather than the endpoint of a branch, may have an effect for finite-size systems.

% ===========================================================
% ===========================================================

Lastly, we mention that the phase boundaries meet at the point $(a_\text{c},1)$.
Accurate determination of $a_\text{c}$ is difficult, even for SAWs \cite{Bradly2018}.
In principle, we can estimate $a_\text{c}$ by restricting the analysis to the fixed value \smash{$y=1$}.
A simple means of locating the adsorption transition is to find the maximum of the quantity
\begin{equation}
    \Gamma_n(a) = \frac{d\log u_n^{(m)}}{dT} =
	(\log a)^2
	\frac{\langle m^2 \rangle - \langle m \rangle^2}{\langle m \rangle},
    \label{eq:LogDerivative}
\end{equation}
at different $n$.
The positions $a_{n,\text{peak}}$ of the peaks are extrapolated to infinite lengths assuming a simple scaling law, \smash{$a_\text{c}-a_{n,\text{peak}} \sim n^{\Delta}$}.
Using this method we estimate the critical temperature \smash{$a_\text{c}=1.34(2)$}, which is consistent with our recent estimate for SAWs of \smash{$a_\text{c}=1.329(2)$} \cite{Bradly2018}.
However, we know from the study of the adsorption of SAWs that this method is too simplistic for accurate analysis. 
The main impediment in this case are the large finite size effects.
Better knowledge of any multicritical scaling and data for much larger branch lengths would be needed to improve the study of this point.

%%%%%%%%%%%%%%%%%%%%%%%%%%%%%%%%%%%%%%%%%%%%%%%%%%%%%%%%%%%%%
%%%%%%%%%%%%%%%%%%%%%%%%%%%%%%%%%%%%%%%%%%%%%%%%%%%%%%%%%%%%%
\section{Conclusion}
\label{sec:Conc}

We have simulated uniform 3-star lattice polymers on the simple cubic lattice up to branch length \smash{$n=128$} subject to a pulling force and adsorption to an interacting surface.
We verify that this system has four phases: free, fully adsorbed, ballistic and a mixed phase.
The transition from the free phase to the ballistic and adsorbed phases is second-order and occurs at the expected critical points \smash{$y_c=1$} and \smash{$a_c=1.34(2)$}, respectively.
The transitions from the ballistic and adsorbed phases to the intermediate mixed phase are shown to be first-order and the location of the boundaries in the asymptotic regime of large $a$ and $y$ agrees with theory.

By altering the set of moves available to each branch the presence of the surface can be simulated without having to predetermine the location of the central node with respect to the surface.
This allows a growth algorithm like flatPERM to be used to access all kinds of configurations of the polymer and thus map out the phase space effectively.
The simulation method means we are studying a slightly different model whereby the star is tethered to the surface at an arbitrary interior point, but we have shown that this does not have any real effect on the phase diagram.

Having demonstrated the case of 3-stars on the cubic lattice, this numerical methodology can now be applied to other related problems for which some theoretical results exist.
A particularly interesting question is how do pulled $f$-stars behave in two-dimensions, for example on the square lattice, where the adsorption of one branch is screened by another, preventing adsorption of the entire $f$-star to the surface.
The theoretical method discussed in \sref{sec:FreeEnergy} fails for two dimensions, but other methods, such as exact enumeration \cite{Guttmann2014}, can proved accurate knowledge of the free energies $\kappa(a)$ and $\lambda(y)$ for SAWs on the square lattice.
The phase boundaries for two dimensions could thus be calculated using other numerical techniques and compared to Monte Carlo results.

Other questions include the adsorption transition of $f$-stars without the pulling force, how the transitions depends on the number of branches.
One expects that the adsorption-mixed boundary does not depend on $f$ since in all cases it involves a single branch being desorbed from the surface, whereas the ballistic-mixed transition involves a further $f-2$ arms being desorbed from the surface.
Away from $f$-stars, SAWs with the pulling force applied at an interior vertices instead of the endpoint is another system that is expected to have an additional phase \cite{Rensburg2017}.
These questions will be the subject of future work.

%%%%%%%%%%%%%%%%%%%%%%%%%%%%%%%%%%%%%%%%%%%%%%%%%%%%%%%%%%%%%
%%%%%%%%%%%%%%%%%%%%%%%%%%%%%%%%%%%%%%%%%%%%%%%%%%%%%%%%%%%%%
\begin{acknowledgments}
C.~B.~ is grateful to Stu Whittington of the University of Toronto and E.~J.~Janse van Rensburg of the York University, Toronto, for fruitful discussions on parts of this work.
C.~B.~thanks York University, Toronto for hosting while part of this work was carried out, as well as funding from the University of Melbourne's ECR Global Mobility Award.
Financial support from the Australian Research Council via its Discovery Projects scheme (DP160103562) is gratefully acknowledged by the authors.
\end{acknowledgments}

%%%%%%%%%%%%%%%%%%%%%%%%%%%%%%%%%%%%%%%%%%%%%%%%%%%%%%%%%%%%%
%%%%%%%%%%%%%%%%%%%%%%%%%%%%%%%%%%%%%%%%%%%%%%%%%%%%%%%%%%%%%
%

\end{document}